\documentclass[lettersize,journal]{IEEEtran}

\usepackage{amsmath,amsfonts}
\usepackage[T1]{fontenc}
\usepackage{textcomp}
\usepackage{algorithm}
\usepackage{algorithmic}
\usepackage{array}
\usepackage{stfloats}
\usepackage{url}
\usepackage{graphicx}
\usepackage{caption}
\usepackage{xcolor}
\usepackage{colortbl}
\usepackage{tabularx}
\usepackage{booktabs}
\usepackage{makecell}
\usepackage{multirow}
\usepackage{seqsplit}
\usepackage{tikz}
\usetikzlibrary{arrows.meta,positioning,calc,fit,shapes.geometric,decorations.pathreplacing,backgrounds}
\usepackage{pgfplots}
\pgfplotsset{compat=1.17}
\usepackage[hidelinks]{hyperref}
\usepackage{orcidlink}

\makeatletter
\renewenvironment{IEEEbiographynophoto}[1]{%
\if@IEEEbiographyTOCentrynotmade%
\setcounter{IEEEbiography}{-1}%
\refstepcounter{IEEEbiography}%
\addcontentsline{toc}{section}{Biographies}%
\global\@IEEEbiographyTOCentrynotmadefalse%
\fi%
\refstepcounter{IEEEbiography}%
\addcontentsline{toc}{subsection}{#1}%
\normalfont\@IEEEcompsoconly{\sffamily}\footnotesize\interlinepenalty500%
\vskip 1\baselineskip plus 0\baselineskip minus 0\baselineskip%
\parskip=0pt\par%
\noindent\textbf{#1\ }\@IEEEgobbleleadPARNLSP}{\relax\par\normalfont}
\makeatother

\providecommand{\xmark}{\ensuremath{\times}}

\begin{document}

\title{Evidence-Driven LLM Agent for C-to-Synthesizable-C Conversion and Verification}

\author{Zhe~Zhao\,\textsuperscript{*}~\orcidlink{0009-0002-8610-1365},
        Hongbing~Lang\,\textsuperscript{*},
        Zhihan~Xiao~\orcidlink{0009-0003-7082-0792},
        Luke~Ztz~Hu~\orcidlink{0009-0007-9450-1170},
        John~Imoleayo~Adebisi~\orcidlink{0009-0009-0469-8980},
        and~Songping~Mai,~\IEEEmembership{Member,~IEEE}~\orcidlink{0000-0002-0572-9066}
\thanks{Zhe Zhao and Hongbing Lang contributed equally to this work. \textit{(Corresponding author: Songping Mai.)}}
\thanks{Zhe Zhao, Zhihan Xiao, Luke Ztz Hu, and Songping Mai are with the Shenzhen International Graduate School, Tsinghua University, Shenzhen 518055, China (e-mail: zhaoz24@mails.tsinghua.edu.cn; mai.songping@sz.tsinghua.edu.cn).}
\thanks{Hongbing Lang is with Shenzhen Belon Technology Co., Ltd., Shenzhen 518055, China (e-mail: hblang@belon.cn).}
\thanks{This work was supported by the Shenzhen Science and Technology Innovation Bureau under Grant ZDCY20250901110707008.}
}

\markboth{IEEE Transactions on Computer-Aided Design of Integrated Circuits and Systems,~Vol.~XX, No.~X, Month~Year}%
{Zhao \MakeLowercase{\textit{et al.}}: LLM Agent for C-to-Synthesizable-C Conversion and Verification}

\maketitle

\bstctlcite{IEEEexample:BSTcontrol}

\begin{abstract}
Software-compilable C programs routinely fail to complete the four-stage pipeline of a high-level synthesis (HLS) toolchain---compilation, C simulation (CSim), synthesis, and C/RTL co-simulation (CoSim)---because HLS accepts only a synthesizable subset of C (HLS-C). Yet most existing large language model (LLM) systems built for HLS code repair only cover the early pipeline stages and feed raw tool logs directly to the model, yielding brittle and hard-to-reproduce fixes. We formulate C-to-HLS-C conversion as a closed-loop generation--verification--diagnosis--repair problem on an HLS tool (Xilinx Vitis), contributing three components: an end-to-end workflow of cooperating agents closed by the four-stage verifier under strict evidence isolation; a Progressive Mismatch Localization Chain (PMLC) that localizes CSim/CoSim mismatches through log normalization, AST backward slicing, and dual-trace instrumentation; and a typed-query, two-stage evidence RAG backed by a self-evolving, family-routed repair-card pool. Experimental results show that the proposed workflow substantially outperforms all comparable state-of-the-art models.
\end{abstract}


\section{Introduction}
\label{sec:intro}

\begin{figure}[!htbp]
\centering
\includegraphics[width=\columnwidth]{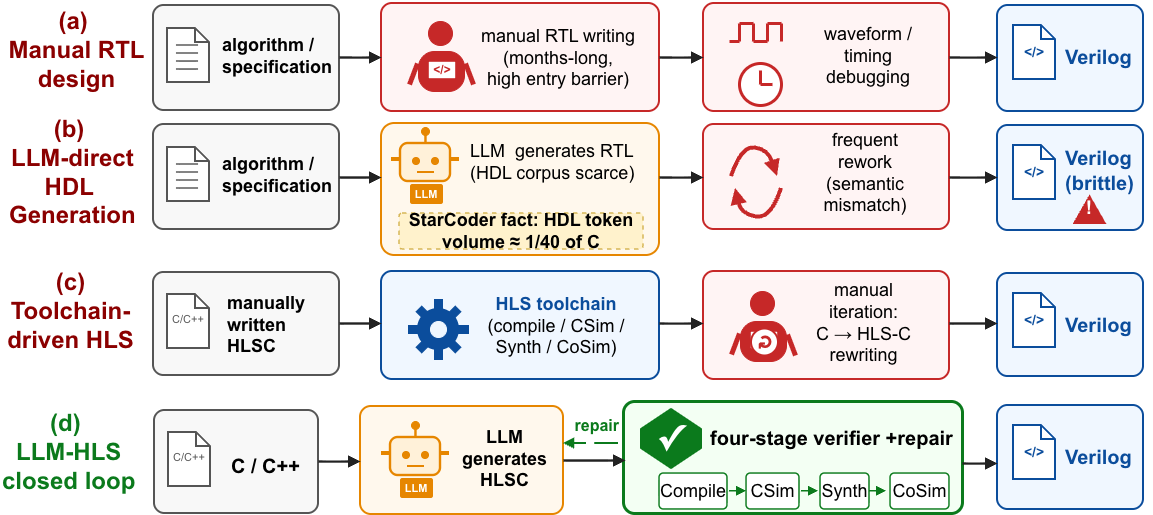}
\caption{Four hardware-design paradigms compared. (a) hand-written RTL is mature but slow and hard to debug; (b) direct LLM-to-HDL is constrained by the scarcity of HDL training corpora; (c) the standard HLS toolchain accepts C/C++ as a hardware entry point, yet C-to-HLS-C rewriting is still largely manual; (d) our work automates C-to-HLS-C generation with an LLM and drives evidence-grounded repair through a four-stage HLS verifier in closed loop.}
\label{fig:design-paradigms}
\end{figure}

\IEEEPARstart{H}{igh}-level synthesis (HLS) raises the abstraction of hardware design from RTL to the C/C++ level, delegating synthesizable RTL generation to the toolchain~\cite{nane2015survey,lahti2018we,cong2022fpga}. Mainstream HLS tools, however, accept only a restricted subset that we call high-level synthesizable C (HLS-C): arbitrary recursion, dynamic memory, STL containers, irregular pointers, and several interface idioms typically lie outside it, so a program that compiles cleanly under GCC often fails the four-stage HLS pipeline of compile, C simulation (CSim), synthesis, and C/RTL co-simulation (CoSim). A natural alternative---letting a large language model (LLM) emit HDL directly---is constrained by the scarcity of HDL training corpora (StarCoder reports HDL at roughly $1/40$ of the C volume~\cite{li2023starcoder}) and tends to yield high error rates and uneven quality~\cite{thakur2024verigen,liu2023verilogeval,lu2024rtllm,liu2024rtlcoder,tsai2024rtlfixer}. A more practical compromise lets the LLM synthesize HLS-C that a mature toolchain already accepts and then lowers it into functionally consistent Verilog, combining the semantic reasoning of LLMs with the engineering-grade constraint solving of HLS. Fig.~\ref{fig:design-paradigms} contrasts the four paradigms; this paper focuses on paradigm (d).

A growing body of LLM-for-HLS work spans C-to-HLS-C translation, benchmarking, debugging prompts, and agentic workflows~\cite{collini2025c2hlsc,zou2025hlstrans,li2025chathls,abi2025hls} (compared in Section~\ref{sec:related}), yet three gaps remain. First, most stop at compile or CSim, while a deployable HLS-C must also pass synthesis and CoSim, the latter guaranteeing cycle-level equivalence between the generated Verilog and the original C. Second, on a CSim/CoSim mismatch the raw log reports only which cycle and output failed, never the root cause; feeding the entire log back generally yields low repair success and unbounded iterations. Third, the few RAG-based systems treat the knowledge base as a static rule book, neither aligned with toolchain failure families nor evolved over the system's own experiments. The underlying difficulty is intrinsic: handing GCC-clean C to Vitis HLS can trigger coupled rewrites cascading across compile $\rightarrow$ CSim $\rightarrow$ synthesis $\rightarrow$ CoSim along five dimensions---interface/signature, memory/pointers, non-synthesizable control flow, bitwidth/numerics, and timing/cycle alignment---so that single-point patches rarely suffice (a concrete walkthrough appears in Section~\ref{subsec:method-mismatch}).

Building on these observations, this paper formulates LLM-driven conversion of standard C into HLS-C---the C-to-HLS-C task---as a closed-loop generation--diagnosis--repair problem governed by the four-stage Vitis HLS verifier rather than as one-shot translation. The workflow is distinguished by three principled choices---evidence-grounded prompt isolation under hard discipline, progressive evidence localization for CSim/CoSim mismatches, and a self-evolving MoE RAG drawn from audited experiments---each anchoring one contribution below. All baselines are reproduced under gpt-5-mini and Vitis HLS 2022.1 with a shared manifest and repair budget K=3 (Section~\ref{subsec:eval-setup}).

\begin{itemize}
    \item An end-to-end closed-loop workflow for C-to-HLS-C-to-Verilog. We target Vitis HLS with the four-stage all-pass of compile/CSim/synthesis/CoSim as the sole headline metric; the deliverable is functionally consistent Verilog, not a snippet that merely compiles. Evidence admitted into the LLM context is strictly limited to a compact task brief, the current failing code, the failing stage, the raw error excerpt, and a local code window, while reference HLS solutions, testbench labels, and baseline private outputs stay audit-only (Section~\ref{subsec:problem-evidence-layer}). On CFull107 (107 designs, gpt-5-mini, K=3), the workflow reaches 94/107 (87.85\%) four-stage all-pass---a +53.27\% absolute gain over the strongest baseline (37/107), with the all-pass rate reaching 2.5x that baseline under the protocol reported here (Table~\ref{tab:cfull107}).
    \item PMLC: a three-layer mismatch evidence chain. For CSim/CoSim mismatches---the hardest failure family---a Progressive Mismatch Localization Chain composes log normalization, AST backward slicing, and selective instrumentation with dual-trace alignment, packaging them with the failing code and a must-preserve contract into a repair brief that shrinks the search space to a few suspect lines, anchoring repair to checkable runtime observations rather than black-box guessing. On the mismatch-dense subset $\mathcal{M}$ ($|\mathcal{M}|=24$) it lifts four-stage all-pass from the strongest baseline's 20.83\% to 70.83\% (17/24).
    \item An MoE-style two-stage evidence RAG (MoE RAG) with a self-evolving card pool. A repair-only RAG hard-routes typed queries to family-specific card subsets and then runs structured exact matching plus GraphRAG-style cross-case linking, returning hints only when stage, family, and symbol match exactly and otherwise short-circuiting to an empty hit. The pool grows from human-audited repair chains, forming a self-evolving loop of run-experiment $\rightarrow$ extract-evidence $\rightarrow$ human-audit $\rightarrow$ enrich-knowledge $\rightarrow$ improve-retrieval, and reduces the average repair-iteration count from $\approx 2.8$ for the baselines to 1.577.
\end{itemize}

The remainder of the paper is organized as follows. Sections~\ref{sec:related} and~\ref{sec:problem} review related work and formalize the problem under the four-stage verifier; Section~\ref{sec:method} unfolds the closed-loop workflow, PMLC, and the MoE RAG; Section~\ref{sec:evaluation} reports the CFull107 main table, cross-model robustness, and component ablations; Section~\ref{sec:conclusion} concludes.


\section{Related Work}
\label{sec:related}

This section organizes the literature along two threads---generation--translation--repair, and RAG with multi-agent---that together respond to the three shared gaps highlighted in Section~\ref{sec:intro} (stopping at compile/CSim, feeding raw logs back unchanged, and treating the knowledge base as a static rule book). For each thread we name representative works and locate the differences from this paper, with emphasis on axes related to system-level closed-loop and repair fidelity; a six-axis method portrait is summarized in Table~\ref{tab:rw-axis-matrix}.

Classical HLS automation has established the C/C++-to-RTL pipeline---scheduling, pipelining, pragma insertion, memory partitioning, and design-space exploration~\cite{nane2015survey,lahti2018we,cong2022fpga,schafer2019high,ferrandi2021bambu,josipovic2021c,sohrabizadeh2022autodse}---with standard benchmarks such as CHStone~\cite{hara2008chstone}, MachSuite~\cite{reagen2014machsuite}, Rosetta~\cite{zhou2018rosetta}, and HLSyn~\cite{bai2023towards}. These efforts hold that engineering value requires synthesis through a commercial toolchain followed by co-simulation; we adopt the same position and confine our headline metric to the four-stage all-pass under Vitis HLS.

LLM-for-HLS generation, translation, and repair forms the closest comparison to this paper. On the generation/translation side, C2HLSC~\cite{collini2025c2hlsc} provides a first feasibility study of LLM-driven C-to-HLS-C; HLSTrans~\cite{zou2025hlstrans} couples MCTS-style search with DSE to assemble a 23K-variant supervised corpus; ChatHLS~\cite{li2025chathls} performs multi-agent repair and optimization with SFT-tuned Llama3-8B agents (implemented via LoRA); HLSPilot~\cite{xiong2024hlspilot} chains profiling, rewrite, and DSE; HLS-Eval~\cite{abi2025hls} is an evaluation suite; SAGE-HLS~\cite{khan2025sage} and SynthAI~\cite{sheikholeslam2024synthai} represent syntax-aware fine-tuning and multi-agent forward design, respectively. On the repair side, the auto-repair line~\cite{xu2024automated,xu2026hlsrewriter} typically drives a single round of fixing from compile-log signals (HLSRewriter is the journal-extended version), HLSDebugger~\cite{wang2025hlsdebugger} performs supervised localization on a pre-labeled bug pool, and CorrectHDL~\cite{xu2025correcthdl} injects HLS-side golden HDL into the prompt for differential repair. Our closed-loop repair is structurally aligned with the LLM-APR thread~\cite{monperrus2018automatic,xia2023automated,joshi2023repair,bouzenia2025repairagent}; complementary LLM-for-RTL studies~\cite{thakur2024verigen,liu2023verilogeval,lu2024rtllm,liu2024rtlcoder,tsai2024rtlfixer} often expose the failure modes of generating HDL directly.

RAG and multi-agent systems form a second thread. VeriMoA~\cite{ping2025verimoa} aggregates parallel proposers under a mixture-of-agents framework; DeepV~\cite{ibnat2025deepv} offers a model-agnostic RAG framework on the Verilog side; LIFT~\cite{prakriya2025lift} and TimelyHLS~\cite{mashnoor2025timelyhls} target pragma insertion and timing-aware optimization, respectively; AgentTTS~\cite{wang2026agenttts} discusses test-time compute budgeting from an agent perspective; the limits of reasoning models in hardware design are reported by~\cite{collini2025can}. General-purpose RAG~\cite{lewis2020retrieval,asai2024self,edge2024local}, agent frameworks~\cite{yao2022react,madaan2023self,wu2023autogen}, and EDA-question-answering RAG~\cite{shi2024ask,kaintura2024orassistant} provide conceptual reference points. The knowledge bases in these works are typically either static high-quality corpora or tool manuals; the principled difference of our approach is that the card pool is distilled from the system's own four-stage verification experiments and admitted only after human audit, with retrieval routed by failure family through a lightweight classifier rather than by similarity alone.

Method portrait and differentiation are summarized in Table~\ref{tab:rw-axis-matrix}, which characterizes representative works along six attributes: tool, weight training, cycle-level mismatch localization, four-stage all-pass, independence from reference HLS or bug labels, and verifier closed-loop repair. Among them, ours is one of the few simultaneously satisfying the comprehensive triple of end-to-end closure to CoSim, no reliance on reference HLS or bug labels, and a knowledge base distilled from the system's own experiments.

\begin{table*}[!htbp]
\caption{Method Portrait of Prior LLM-for-HLS Works on Six Capability Axes (\checkmark{} supported, \xmark{} not supported, $-$ out of scope).}
\label{tab:rw-axis-matrix}
\centering
\footnotesize
\setlength{\tabcolsep}{3.8pt}
\renewcommand{\arraystretch}{1.22}
\begin{tabularx}{\textwidth}{l c c c c c c >{\centering\arraybackslash}p{5.5cm}}
\toprule
\textbf{Method [ref]} & \textbf{Tool} & \textbf{Train} & \makecell[c]{\textbf{Cycle-level}\\\textbf{Mism. Loc.}} & \makecell[c]{\textbf{4-Stage}\\\textbf{All-Pass}} & \makecell[c]{\textbf{No Ref-HLS}\\\textbf{/ Bug Label}} & \makecell[c]{\textbf{Verifier}\\\textbf{Closed-Loop}} & \textbf{Notes} \\
\midrule
ChatHLS \cite{li2025chathls}             & Vitis     & SFT  & \xmark      & partial   & \xmark    & partial   & Multi-agent repair + QoR \\
HLSDebugger \cite{wang2025hlsdebugger}     & in-house  & SFT  & \xmark      & \xmark    & \xmark    & \xmark    & Supervised top-$k$ ranking \\
C2HLSC \cite{collini2025c2hlsc}               & Catapult  & none & \xmark      & \xmark    & \checkmark & \xmark    & One-shot prompt + raw log \\
HLSPilot \cite{xiong2024hlspilot}           & Vitis     & none & \xmark      & \xmark    & \checkmark & \xmark    & Profiling$\to$DSE \\
HLS-Eval \cite{abi2025hls}            & Vitis     & none & \xmark      & partial   & \checkmark & \xmark    & Static evaluation only \\
Auto-repair \cite{xu2024automated} & Catapult  & none & \xmark      & \xmark    & \checkmark & \checkmark & Compile-log single round \\
CorrectHDL \cite{xu2025correcthdl}       & Catapult  & none & \xmark      & partial   & \xmark    & partial   & Differential vs.\ golden HDL \\
VeriMoA \cite{ping2025verimoa}             & iverilog  & none & $-$         & $-$       & \checkmark & partial   & RTL path; multi-proposer \\
\midrule
\textbf{This work}                     & \textbf{Vitis} & \textbf{none} & \textbf{\checkmark} & \textbf{\checkmark} & \textbf{\checkmark} & \textbf{\checkmark} & \textbf{4-agent loop + PMLC + MoE RAG} \\
\bottomrule
\end{tabularx}
\end{table*}


\section{Problem Formulation}
\label{sec:problem}

This section fixes the formal objects shared by Sections~\ref{sec:method}--\ref{sec:evaluation}---task instances, the four-stage verifier, the prompt-facing/artifact-only evidence split, and the repair recursion---which aim at substantial auditability and serve as the principled basis for the claims that follow; baseline reproduction-fidelity labels appear in Section~\ref{subsec:eval-setup}.

\subsection{Task Instances and the Four-Stage Verifier}
\label{subsec:problem-instance}
\label{subsec:problem-verifier}

For item $i$ in manifest $\mathcal{D}$, the input is $x_i = (c_i, \Lambda_i, q_i)$: the C/C++ program $c_i$, the public HLS testbench/TCL context $\Lambda_i$, and the externally visible top-function and interface contract $q_i$ (must-preserve signature, ports, types, timing). An agent $G_{\theta}$ produces a candidate $y_i^{(0)}$ from a compact brief $b_i = P(c_i, \Lambda_i, q_i)$, where $P$ is lossy yet interface-lossless; the full hard-isolation statement is in Section~\ref{subsec:problem-evidence-layer}.

The four-stage HLS verifier runs the stages in $S=\{\mathrm{compile},\mathrm{csim},\mathrm{synth},\mathrm{cosim}\}$. For each stage $s\in S$, the verifier component $V_s$ returns a Boolean indicator $v_s$ and the raw log $\ell_s$:
\begin{equation}
V_s(y_i) = \big(v_s(y_i),\, \ell_s(y_i)\big),\quad s \in S.
\label{eq:stage}
\end{equation}
Stages run in the order of $S$ and short-circuit on the first failure (unexecuted stages set $v_s = 0$ by convention), giving the four-stage all-pass indicator and earliest failing stage $s^*(y_i) = \arg\min_{s}\{s : v_s(y_i) = 0\}$:
\begin{equation}
\mathrm{AllPass}(y_i) = \prod_{s \in S} \mathbb{1}\!\left[v_s(y_i) = 1\right].
\label{eq:allpass}
\end{equation}

A repair budget $K \in \mathbb{Z}_{>0}$ bounds the loop ($K = 3$ on CFull107; any directly comparable baseline shares the same $K$); with $y_i^{(k)}$ the candidate after round $k$ and $\sigma_i$ the execution trace, the round count follows the stopping rule
\begin{equation}
K_i = \min\bigl\{ k \le K \;\big|\;\, \mathrm{AllPass}(y_i^{(k)}) = 1 \;\lor\; k = K \bigr\}.
\label{eq:stop}
\end{equation}
The success rate is $\mathrm{SR}(\mathcal{D}) = \frac{1}{|\mathcal{D}|}\sum_{i} \mathrm{AllPass}(y_i^{(K_i)})$, and lift over the strongest baseline uses
\begin{equation}
\Delta_{\mathrm{rel}}(a, b) = \frac{\mathrm{SR}_a - \mathrm{SR}_b}{\mathrm{SR}_b},
\label{eq:lift}
\end{equation}
computed only when the two rows share manifest, model, and budget with $\mathrm{SR}_b > 0$; transfer rows that adopt a prior paper's own manifest generally report only the absolute difference, with a caveat.

\subsection{Prompt-Facing Evidence vs.\ Artifact-Only Trace}
\label{subsec:problem-evidence-layer}

Within all persisted payloads $\Omega_i^{(k)}$ at round $k$ (excluding the LLM's hidden state), we split the prompt-facing evidence $E_i^{(k)}$ from the audit-only artifacts $\mathcal{A}_i^{(k)}$:
\begin{equation}
E_i^{(k)} \subseteq \Omega_i^{(k)}, \quad \mathcal{A}_i^{(k)} \subseteq \Omega_i^{(k)}, \quad E_i^{(k)} \cap \mathcal{A}_i^{(k)} = \emptyset.
\label{eq:omega-split}
\end{equation}
$E_i^{(k)}$ admits only five items: the brief $b_i$; the current candidate $y_i^{(k)}$; the failing stage $s^*(y_i^{(k)})$; a truncated first-error excerpt $\widehat{\ell}_{s^*}$ (default $80$ lines); and a local window $W_i^{(k)}$ of radius $8$ around the log-anchored symbols. On the mismatch stage it additionally admits the three layers
\begin{equation}
E_i^{(k),\,\mathrm{mismatch}} = (L_1, L_2, L_3),
\label{eq:l123}
\end{equation}
with $L_1$ output-level (failed outputs and first failing cycle), $L_2$ source-level (a backward slice around the named symbols of $L_1$), and $L_3$ runtime evidence (a value spectrum inside the $L_2$ scope); budgets $\kappa_j$ and schemas appear in Section~\ref{sec:method}. The artifact-only set $\mathcal{A}_i^{(k)}$ holds the full log, the reference HLS $y_i^{\mathrm{ref}}$, testbench labels $\Lambda_i^{\mathrm{label}}$, manual repairs $y_i^{\mathrm{manual}}$, baseline private outputs, and the full trace $\sigma_i$. We impose hard isolation from the leakage channel $L^{\mathrm{leak}} = \{y_i^{\mathrm{ref}}, \Lambda_i^{\mathrm{label}}, y_i^{\mathrm{manual}}, \{y_i^{\mathrm{baseline},m}\}\}$, stated formally as Proposition~2 (Section~\ref{subsec:method-evidence-sufficiency})---typically the principled distinction from baselines depending on reference HLS or pre-labeled bugs. Concretely, the admitted $\Lambda_i$ is only the verification harness---the I/O wrapper, the driver/TCL, the stimulus generator, and the top-function signature---authored from the original C and its specification before any HLS-C exists, so it encodes no HLS-C-specific solution; its golden reference is $c_i$ executed at run time rather than a stored expected-output table, while the stored golden labels $\Lambda_i^{\mathrm{label}}$ stay in $\mathcal{A}_i^{(k)}$. Hence $\Lambda_i$ adds no information about correct behavior beyond what $c_i$ already specifies and merely pins down the must-preserve contract $q_i$; a degenerate ``hardcode-the-outputs'' HLS-C is not constructible from $E_i^{(k)}$ and, even if attempted, would fail synthesis, the interface contract, and cycle-level CoSim over the full stimulus set. Our prompt-facing leakage surface is therefore strictly smaller than that of baselines admitting reference HLS or pre-labeled bugs.

\subsection{Repair Recursion and Failure Analysis}
\label{subsec:problem-repair}

Each round ($k \ge 1$) updates the candidate by
\begin{equation}
y_i^{(k)} = R_{\theta}\!\Big(y_i^{(k-1)},\; A_{\theta}\big(E_i^{(k-1)}\big),\; q_i\Big),
\label{eq:repair}
\end{equation}
where $R_{\theta}$ and $A_{\theta}$ are the LLM repair and failure-analysis operators, $E_i^{(k-1)}$ is the isolated failure evidence under \eqref{eq:omega-split} (mismatch rounds use \eqref{eq:l123}), and $q_i$ is the must-preserve contract. Crucially, $A_{\theta}$ and $R_{\theta}$ never see the full trace $\sigma_i$---only the most recent compressed failure---which generally avoids long-context drift and historical hallucination; cross-round history is retained only in the audit ledger inside $\mathcal{A}_i^{(k)}$. These definitions fix tasks, metrics, and symbol boundaries but do not constitute a convergence proof; the agent implementation and three-layer evidence are detailed in Section~\ref{sec:method}.


\section{Method}
\label{sec:method}

This section instantiates the formal objects of Section~\ref{sec:problem}---task instances, the four-stage verifier, the stopping rule, and the $E_i^{(k)}/\mathcal{A}_i^{(k)}$ split---as concrete operators and data structures, organized as the closed-loop workflow, the prompt contract, PMLC, the MoE RAG, and an evidence-sufficiency argument.

\subsection{Overall Closed-Loop Workflow with Role-Specialized Agents}
\label{subsec:method-overall}

\begin{figure*}[t]
\centering
\includegraphics[width=0.92\textwidth]{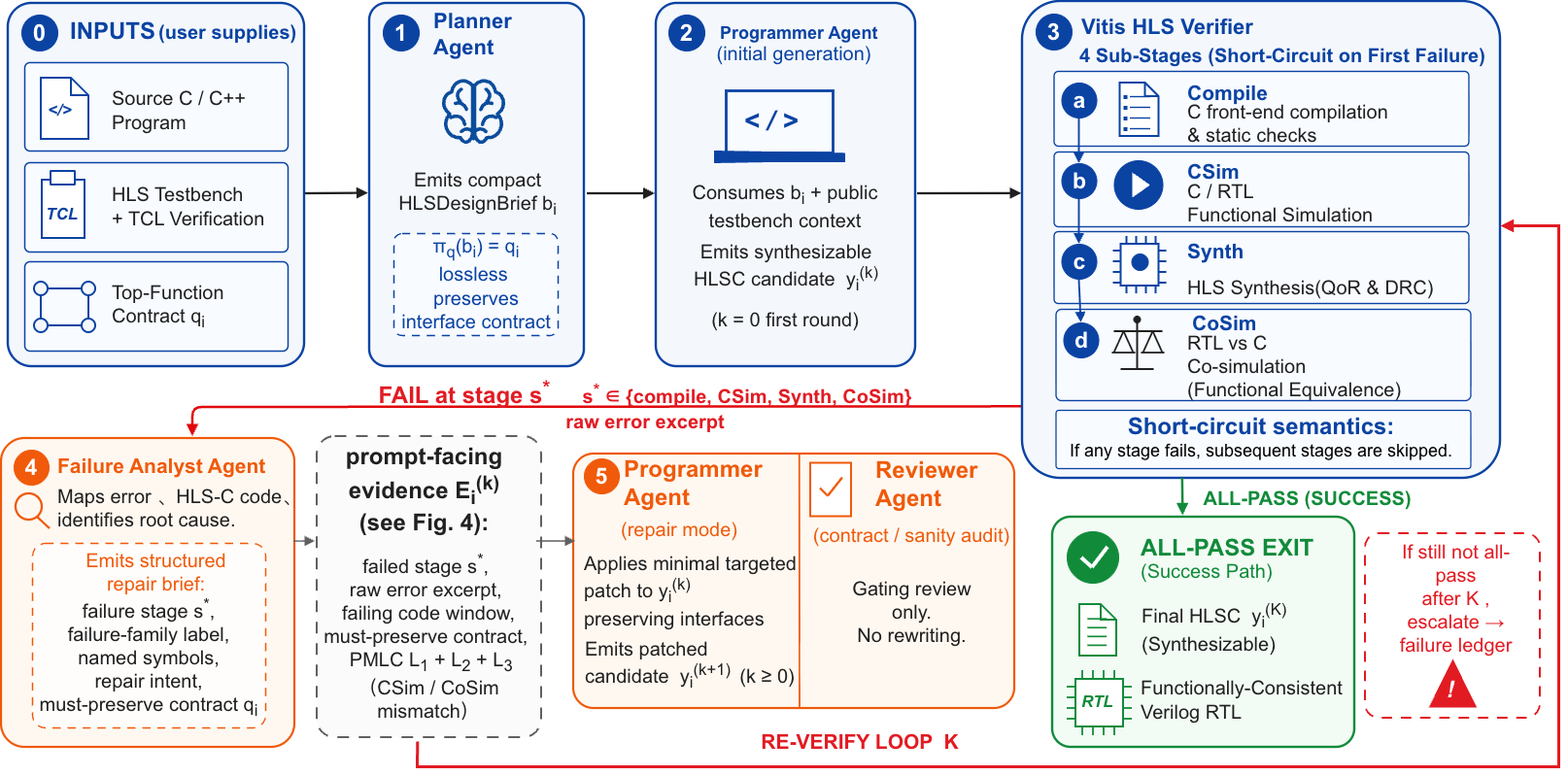}
\caption{C-to-HLS-C closed loop. The planner projects an interface-lossless brief $b_i$ ($\pi_q(b_i)=q_i$), the programmer emits a candidate, and the Vitis HLS verifier short-circuits along compile/CSim/synthesis/CoSim; on a failure at $s^{\ast}$ the failure analyst builds a repair brief from the prompt-facing evidence $E_i^{(k)}$ and the programmer applies a minimal patch, with PMLC layers $L_1{+}L_2{+}L_3$ appended on mismatches.}
\label{fig:overall-framework}
\end{figure*}

As shown in Fig.~\ref{fig:overall-framework}, the system distributes work across four lightweight agents. The planner projects the compact task brief $b_i$ while preserving the interface contract $\pi_q(b_i) = q_i$; the programmer generates the candidate $y_i^{(0)}$ from $b_i$; Vitis HLS verifies in the order compile, CSim, synthesis, CoSim; and on any failing stage the orchestrator invokes the evidence-extraction layer to produce $E_i^{(k)}$, which the failure analyst $A_{\theta}$ turns into a repair brief and the programmer applies through $R_{\theta}$. The reviewer performs only a contract and sanity audit after each patch and does not rewrite programmer outputs. We treat Vitis HLS as the controller of the repair loop rather than a mere pass/fail oracle: a verifier wrapper applies the short-circuit semantics of Section~\ref{subsec:problem-verifier} so that any stage failure halts the rest, while the CoSim stage carries a triple-watchdog (total duration, stdout-silence interval, and subprocess liveness) that distinguishes a long simulation from a deadlock and, when triggered, re-enters the repair loop. The whole process obeys the hard isolation of Section~\ref{subsec:problem-evidence-layer} and the stopping rule of \eqref{eq:stop}.

The role split is intended to reduce prompt contamination: the programmer in repair mode sees only the current failing code and the most recent failure evidence (the stateless update of \eqref{eq:repair}), which generally avoids the hallucination drift caused by accumulating history. In contrast to multi-agent forward-design decompositions~\cite{sheikholeslam2024synthai,ping2025verimoa}, the principled goal of our split is to compress the information reaching the LLM at prompt time and to fix role boundaries, not to add richer intermediate representations.

\subsection{Prompt Contract and Information Isolation}
\label{subsec:method-prompt-contract}

Our prompt design instantiates the $E_i^{(k)}$ admission set of Section~\ref{subsec:problem-evidence-layer} over two prompt classes. The initial generation prompt ($k=0$) carries the source C/C++, a public testbench/TCL summary (the harness and interface only, not the golden expected outputs), the required top-function signature, and the minimal brief $b_i$. The repair prompt ($k \ge 1$) carries the current candidate, the earliest failing stage $s^{\ast}$, the truncated raw tool error $\widehat{\ell}_{s^{\ast}}$, the failure-analyst summary, the patch scope, and the must-preserve contract; the PMLC three layers are attached on the mismatch stage, while the full execution trace $\sigma_i$ and cross-round history are excluded. For the compile and synthesis stages we additionally admit a length-capped project-log hint extracted from the HLS project directory (trailing tool-log lines, TCL conversion fragments, and the testbench interface signature), delivering signals the toolchain has already written without ingesting the full log.

The failure-analyst output is fixed to four structured fields---the truncated raw error excerpt, the failure-family label, the named-symbol set, and the repair intent---plus a must-preserve contract. These serve both the repair prompt and the typed RAG query of Section~\ref{subsec:method-rag}, so that each round poses only three concrete questions: why the current code fails, where the minimal edit should land, and which contracts must not move.

\subsection{PMLC: Progressive Mismatch Localization Chain}
\label{subsec:method-mismatch}

Compile and synthesis errors are typically already addressable from $s^{\ast}$, $\widehat{\ell}_{s^{\ast}}$, and the local window $W_i^{(k)}$. CSim/CoSim mismatches are harder: a 2000-line cycle-by-cycle comparison log tells the LLM only which cycle output failed, not why or where, and replaying it verbatim tends to induce irrelevant edits on loop bounds or unrelated variables. We therefore introduce PMLC (Progressive Mismatch Localization Chain), three evidence layers $L_1,L_2,L_3$ as in~\eqref{eq:l123} that progressively shrink the repair search space from the entire HLS-C to a few suspect lines (monotone shrinkage formalized in Proposition~1).

\begin{figure*}[t]
\centering
\includegraphics[width=0.88\textwidth]{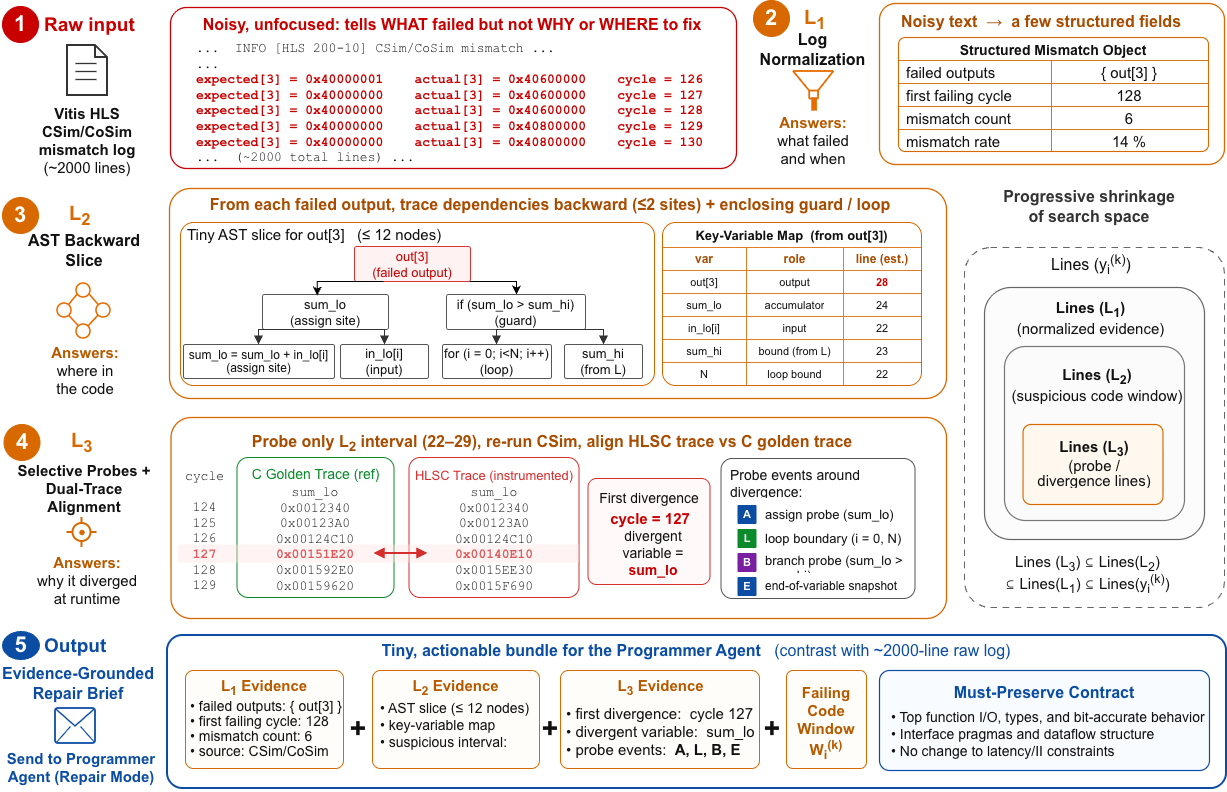}
\caption{PMLC three-layer evidence chain. $L_1$ normalizes the mismatch log into a structured failure object; $L_2$ runs an AST backward slice (depth $\le 2$) to a key-variable map and merged suspect lines; $L_3$ aligns the C golden trace with the HLS-C trace to localize the first-divergence cycle. The layers shrink the search space monotonically, $\mathrm{Lines}(L_3)\subseteq\mathrm{Lines}(L_2)\subseteq\mathrm{Lines}(L_1)\subseteq\mathrm{Lines}(y_i^{(k)})$ (Proposition~1). In-figure numbers illustrate one accumulator-order case from CFull107.}
\label{fig:mismatch-chain}
\end{figure*}

The chain runs as follows (Fig.~\ref{fig:mismatch-chain}). $L_1$ normalizes the simulation log into output-side evidence (matched/total counts, the first failing cycle, the failed-output list, and a few mismatch examples), with a standalone predicate that separates genuine mismatches from compile/tool errors. $L_2$ parses the HLS-C syntax tree and runs an AST backward slice rooted at the failed outputs, yielding a key-variable map of related variables, assignment and control sites, and merged suspect line ranges; recursion depth is bounded so the slice stays precise. $L_3$ then selectively instruments only those sites with lightweight probes (assignments, loop iterations, branch outcomes, variable snapshots) and, when the original C is available as a golden reference, places co-located probes on both C and HLS-C to align the two traces and localize the first-divergence cycle and the diverging variables. A packager bundles the three layers, the failing code window, the project-log summary, the raw error excerpt, and the must-preserve contract into the repair brief. Conceptually $L_2/L_3$ align with spectrum-based fault localization and program slicing, but lift the localization object from program variables to the differential between HLS-C and testbench expectations; we use this analogy under a constrained scope and claim no equivalent guarantees. The principled difference is that PMLC binds mismatch repair to verifiable runtime observations, contrasting with~\cite{collini2025c2hlsc} that replays raw logs verbatim, with~\cite{xu2024automated} driven only by compile-log signals, and with~\cite{wang2025hlsdebugger} that depends on supervised pre-labeled bug pools. A small set of engineering budgets $\kappa_j$ (slice depth, line-merge/expansion thresholds, snapshot retention, and the $80$-line error excerpt with radius-$8$ window) keeps the prompt-facing set $E_i^{(k)}$ decoupled from the input design size (Proposition~1).

\subsection{MoE-Style Two-Stage Evidence RAG with Typed-Query Routing}
\label{subsec:method-rag}

\begin{figure}[t]
\centering
\includegraphics[width=\linewidth]{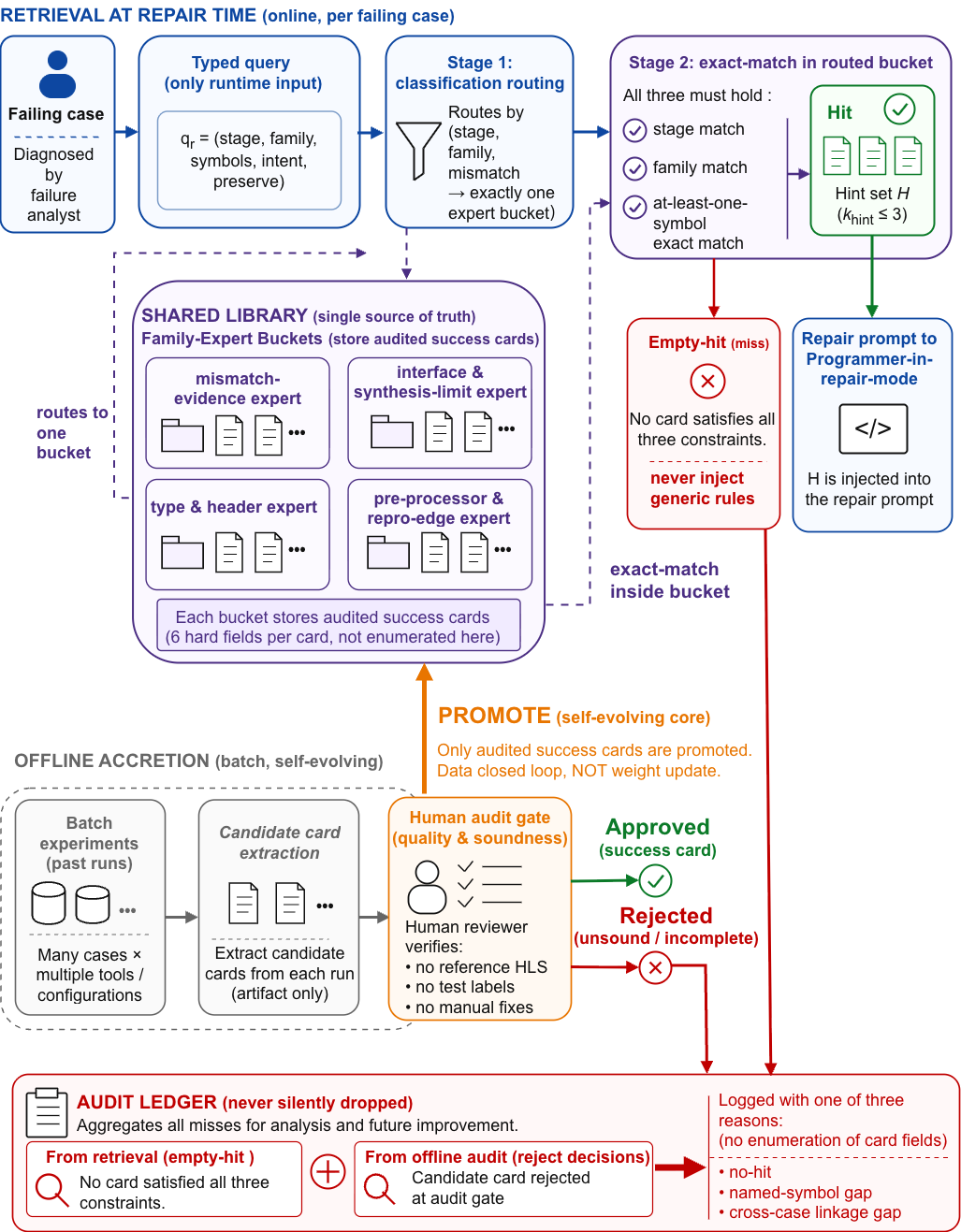}
\caption{Self-evolving MoE RAG. A typed query $q_r$ over the fields stage, family, symbols, intent, and preserve hard-routes to one expert bucket (Stage~1) and returns up to $k_{\rm hint}\!\le\!3$ exact-matched repair cards, or an empty hit on any miss (Stage~2). Offline, human-audited repair chains (no reference HLS, testbench labels, or manual repairs) are promoted into the buckets while audit-failed cases enter the audit ledger, forming a data-side closed loop rather than online weight updates.}
\label{fig:rag-loop}
\end{figure}

RAG is, in our system, neither a replacement for the verifier nor a silent rewriter but an independent repair-only enhancement track, organized as an MoE-style two-stage retrieval (Fig.~\ref{fig:rag-loop}). Stage~1 performs rule-based classification routing from the structured failure-analyst output: by failing stage, failure-family label, and the mismatch-only flag, the failure is hard-routed to a family-specific sub-pool (mismatch-evidence, interface/synthesis-constraint, type/header, or preprocessing/reproduction-boundary experts). Stage~2 runs structured exact matching inside that sub-pool on stage, family, and named-symbol set, plus optional GraphRAG-style cross-case linking; it returns up to $k_{\rm hint}\!\le\!3$ hints only when stage, family, and at least one symbol match exactly, otherwise short-circuiting to an empty hit so that generic rules are never injected. Here MoE-style refers only to hard routing by family label: Stage~1 is a classifier, not a learnable gate, and we do not train expert weights.

The knowledge base holds repair success cards (plus a few baseline-knowledge cards for toolchain compatibility and domain common sense), each carrying six required fields: observed failing stage, validated passing stage, permitted retrieval range, truncated error excerpt, before/after code window, and truncated repair diff. By self-evolving we mean a data-side loop rather than online weight updates: only after a complete failure-to-repair chain is human-audited line by line---confirming no leakage of reference HLS, testbench labels, or manual repairs---is the case promoted into the formal pool, while incomplete cases are recorded in an audit ledger (empty hit, named-symbol gap, cross-case linkage gap) as an auditable record of retrieval blind spots. This forms the loop run-experiment $\rightarrow$ extract-evidence $\rightarrow$ human-audit $\rightarrow$ enrich-knowledge $\rightarrow$ improve-retrieval, keeping retrieval results close to a small set of citable evidence rather than long generic rules.

\subsection{Evidence Sufficiency and Auditable Absence of Leakage}
\label{subsec:method-evidence-sufficiency}

We do not claim a convergence proof for the LLM-driven loop. The two propositions below are not theorems but auditable implementation-level observations on information control and evidence size, verifiable externally through the audit ledger.

\medskip
\noindent\textbf{Proposition 1 (Bounded evidence size and monotone shrinkage).}\\[3pt]
\noindent The size of all three evidence layers is controlled by enumerable budgets $\kappa_j$ and does not grow with input size; moreover the indicated source-code line sets satisfy $\mathrm{Lines}(L_3) \subseteq \mathrm{Lines}(L_2) \subseteq \mathrm{Lines}(L_1) \subseteq \mathrm{Lines}(y)$, shrinking monotonically. These are finite-projection properties of $E_i^{(k)}$, not an existence proof; the empirical support is the full50 localization ablation (Section~\ref{subsec:eval-main}), a controlled diagnostic set of 50 mismatch cases with known ground-truth fault locations, kept separate from CFull107 and used only to score layer-wise accuracy, on which $L_1/L_2$ reach 50/50 and $L_3$ dual-trace first-divergence localization reaches 45/50.

\medskip
\noindent\textbf{Proposition 2 (Auditable absence of reference and label leakage).}\\[3pt]
\noindent Let $L^{\mathrm{leak}} := \{y_i^{\mathrm{ref}}, \Lambda_i^{\mathrm{label}}, y_i^{\mathrm{manual}}, \{y_i^{\mathrm{baseline},m}\}\}$. Our implementation satisfies
\begin{equation}
\forall\,k \ge 0,\; L^{\mathrm{leak}} \cap E_i^{(k)} = \emptyset,
\label{eq:no-leak}
\end{equation}
which can be audited in both the prompt-construction code and the audit ledger.

\begin{figure}[t]
\centering
\includegraphics[width=\columnwidth]{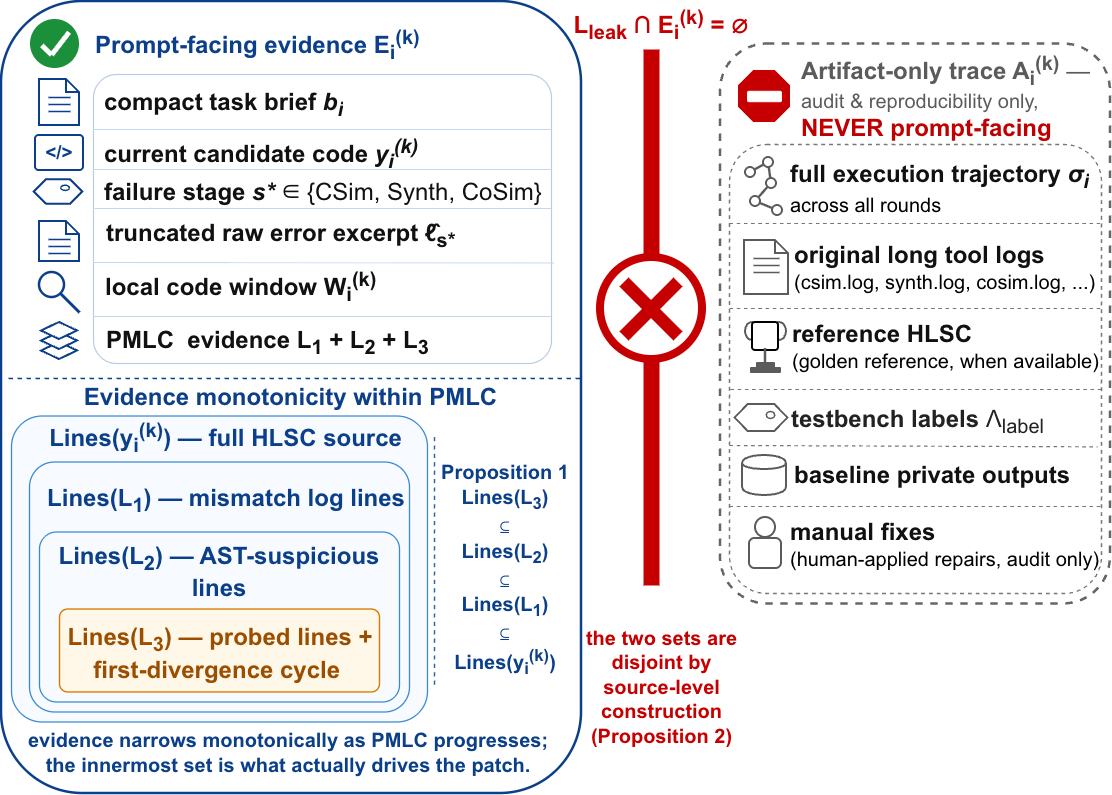}
\caption{Information stratification between prompt-facing ($E_i^{(k)}$) and artifact-only ($\mathcal{A}_i^{(k)}$) (Section~\ref{subsec:problem-evidence-layer}) and the visual locations of Propositions~1 and~2. The forbidden injection in the middle is the constraint of~\eqref{eq:no-leak}, while the three nested layers on the left depict the monotone shrinkage of $\mathrm{Lines}(L_j)$ in Proposition~1.}
\label{fig:info-boundary}
\end{figure}

\noindent The input schemas of the four agents and the RAG retriever contain none of the four leakage channels; the initial and repair prompts are built from two non-overlapping schemas, and cross-round failure information is retained only in the audit ledger inside $\mathcal{A}_i^{(k)}$, so the condition is statically checkable. This is the substantial difference from~\cite{xu2025correcthdl}, which injects golden HDL into the prompt, and from~\cite{wang2025hlsdebugger}, which uses supervised pre-labeled bug pools; it is also the method-level carrier of the hard isolation of Section~\ref{subsec:problem-evidence-layer}. Fig.~\ref{fig:info-boundary} visualizes the partition and locates Propositions~1 and~2.

\section{Experimental Evaluation}
\label{sec:evaluation}

We organize the experiments along three concentric layers: (i) an end-to-end comparison against LLM-driven baselines on CFull107 (Section~\ref{subsec:eval-main}); (ii) a decomposition of workflow gain versus model gain through cross-model and zero-shot rows (Section~\ref{subsec:eval-models}); and (iii) a transferability check on two public HLS benchmarks (Section~\ref{subsec:eval-transfer}). PMLC and RAG ablations are embedded at the end of Section~\ref{subsec:eval-main}.

\subsection{Experimental Setup}
\label{subsec:eval-setup}

\emph{Benchmark.}\ CFull107 is our benchmark of 107 distinct C/HLS-C design pairs. We build it from several established HLS datasets and benchmarks (e.g., MachSuite, PP4FPGA, ForgeBench, and Chrysalis-HLS-style corpora): each pair comprises a standard C program, a reference HLS-C implementation, and an automatically generated, coverage-driven testbench whose golden reference is the original C executed at run time (rather than a stored expected-output table). Construction proceeds in three steps: (i) parameterized augmentation (array dimensions and loop-unrolling factors) for structural diversity; (ii) dual-consistency filtering that retains only pairs satisfying both functional equivalence and coverage adequacy; and (iii) de-duplication of templated and parameterized variants, leaving 107 functionally independent designs that span five semantic subsets---pointer/memory, loop-heavy, array-access, crypto/DSP, and control-heavy. During conversion the agent receives only the harness and the top-function signature, never the golden outputs.

\emph{Models.}\ The main configuration uses gpt-5-mini with temperature=0 and deterministic decoding; the cross-model rows additionally run gpt-4o-mini, DeepSeek~v4-flash, and DeepSeek~v4-pro under the same workflow. All baselines are reproduced with gpt-5-mini and the official prompts of each method.

\emph{Repair budget and headline metric.}\ We adopt pass@1 evaluation (a single sample per case, no $n$-best reranking) together with at most K=3 verifier-driven repair iterations: pass@1 is the within-round sampling protocol, while K=3 is the cross-round iteration cap, and the two are orthogonal. The headline metric is four-stage all-pass (compile, CSim, synthesis, and CoSim all passing); auxiliary diagnostics include per-stage pass rates, the average repair-iteration count, and per-case token / wall-clock cost. Synthesis and co-simulation use Vitis HLS 2022.1.

\emph{Baselines.}\ We select three LLM-driven HLS workflows: C2HLSC~\cite{collini2025c2hlsc}, HLSTrans~\cite{zou2025hlstrans}, and ChatHLS~\cite{li2025chathls}. C2HLSC is reproduced at the method level because no executable inference repository is available; HLSTrans and ChatHLS are re-run under their official prompts with gpt-5-mini and Vitis HLS 2022.1.

\subsection{CFull107 Main Comparison}
\label{subsec:eval-main}

\begin{table*}[!htbp]
\caption{CFull107 Main Results under Shared Protocol ($|\mathcal{D}|{=}107$, Vitis HLS 2022.1, pass@1, K=3).}
\label{tab:cfull107}
\centering
\footnotesize
\setlength{\tabcolsep}{5pt}
\renewcommand{\arraystretch}{1.20}
\begin{tabularx}{\textwidth}{l l c c c >{\centering\arraybackslash}X c c}
\toprule
\textbf{Method} & \textbf{Model} & \textbf{Compile} & \textbf{CSim} & \textbf{Synth} & \textbf{All-Pass} & \textbf{Tok/Case$^{*}$} & \textbf{Time/Case} \\
\midrule
Ours+RAG                       & gpt-5-mini          & 95.33\% & 95.33\% & 89.72\% & \underline{87.85\%} & 18.8K & 140s \\
Ours+RAG                       & gpt-4o-mini         & 84.11\% & 84.11\% & 77.57\% & 74.77\%             & 16.5K & 121s \\
Ours+RAG                       & DeepSeek v4-flash   & 95.33\% & 93.46\% & 93.46\% & 91.59\%             & 19.2K & 131s \\
Ours+RAG                       & DeepSeek v4-pro     & 95.33\% & 95.33\% & 95.33\% & 95.33\%             & 21.5K & 165s \\
\midrule
C2HLSC \cite{collini2025c2hlsc}       & gpt-5-mini  & 51.40\%          & 51.40\%          & 43.93\%          & 34.58\%                        & 14.5K         & 137s         \\
HLSTrans \cite{zou2025hlstrans}   & gpt-5-mini  & 42.99\%          & 42.99\%          & 38.32\%          & 29.91\%                        & 15.6K         & 177s         \\
ChatHLS \cite{li2025chathls}     & gpt-5-mini  & 44.86\%          & 44.86\%          & 38.32\%          & 28.97\%                        & 60.6K         & 335s         \\
\bottomrule
\end{tabularx}

\vspace{3pt}
\par\raggedright\footnotesize $^{*}$Tok/Case reflects each method's original prompting strategy. The gpt-5-mini row is the protocol-aligned headline; the remaining three Ours rows show cross-model robustness.\par
\end{table*}

\emph{Headline result.}\ On CFull107 (107 designs, gpt-5-mini, K=3), our workflow reaches a four-stage all-pass of 94/107 (87.85\%). Against the strongest LLM baseline C2HLSC at 37/107 (34.58\%), this corresponds to an absolute gap of +53.27\% and, by~\eqref{eq:lift}, an all-pass rate that is approximately 2.5x the baseline (Table~\ref{tab:cfull107}). Under the protocol reported here, the gain is robust: the average repair-iteration count drops from 2.76--2.89 for the baselines to 1.577, with 35.51\% (38/107) of designs clearing all four stages on the first round, while baselines typically exhaust the K=3 budget across the board.

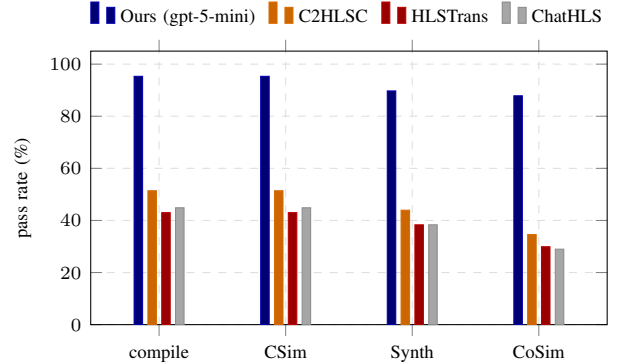
\begin{figure}[t]
\centering
\begin{tikzpicture}
\begin{axis}[
  width=0.95\columnwidth,
  height=52mm,
  ybar,
  bar width=3.2pt,
  enlarge x limits=0.18,
  ymin=0, ymax=105,
  ylabel={pass rate (\%)},
  symbolic x coords={compile, CSim, Synth, CoSim},
  xtick=data,
  ytick={0,20,40,60,80,100},
  tick label style={font=\scriptsize},
  label style={font=\scriptsize},
  legend style={
      font=\scriptsize, at={(0.5,1.05)}, anchor=south,
      legend columns=4, /tikz/every even column/.append style={column sep=3pt},
      draw=none
  },
  grid=major,
  grid style={dashed, gray!25},
]
\addplot+[fill=blue!45!black, draw=blue!70!black] coordinates {
  (compile,95.33) (CSim,95.33) (Synth,89.72) (CoSim,87.85)};
\addplot+[fill=orange!80!black, draw=orange!85!black] coordinates {
  (compile,51.40) (CSim,51.40) (Synth,43.93) (CoSim,34.58)};
\addplot+[fill=red!60!black, draw=red!75!black] coordinates {
  (compile,42.99) (CSim,42.99) (Synth,38.32) (CoSim,29.91)};
\addplot+[fill=gray!70, draw=gray!90] coordinates {
  (compile,44.86) (CSim,44.86) (Synth,38.32) (CoSim,28.97)};
\legend{Ours (gpt-5-mini), C2HLSC, HLSTrans, ChatHLS}
\end{axis}
\end{tikzpicture}
\caption{Per-stage pass rates on CFull107. Ours stays at $\geq 87\%$ across all four stages, while the three LLM baselines visibly attrit at the compile stage (42.99\%--51.40\%) and cannot recover at later stages.}
\label{fig:eval-attrition}
\end{figure}

\emph{Per-stage attrition.}\ Fig.~\ref{fig:eval-attrition} reports per-stage pass rates. Ours stays at 87.85\%--95.33\% across all four stages, with the largest drop at CSim$\to$synthesis (non-synthesizable array/interface idioms), whereas the three baselines already sit at 42.99\%--51.40\% at compile and cannot recover. The compile and CSim pass counts coincide within each row because compile/CSim failures are predominantly C-level semantic errors that the reasoning backbone repairs reliably inside K=3; genuine attrition thus concentrates at the HLS-specific synthesis and CoSim stages. The zero-shot rows of Table~\ref{tab:zeroshot}, produced in a single shot, accordingly show a visible compile$\to$CSim drop.

\emph{Mismatch-dense subset.}\ We define the mismatch-dense subset $\mathcal{M}\subseteq$ CFull107 ($|\mathcal{M}|=24$) by an objective, method-agnostic criterion on the design, not on which method wins: a design enters $\mathcal{M}$ if it clears compile and synthesis but exposes a cycle-level mismatch (``CSim pass, CoSim fail'' or an analogous CSim numerical deviation) under at least one configuration. This isolates the regime where the bottleneck is genuinely the mismatch-repair path rather than a front-stage blocker; it depends only on whether a design ever reaches the mismatch regime, and all methods are scored on the identical 24 designs, so the comparison stays symmetric. A purely source-feature definition would be less faithful, since whether a mismatch is exposed is a runtime property of co-simulation. On $\mathcal{M}$, ours reaches 17/24 (70.83\%), versus C2HLSC 5/24 (20.83\%), ChatHLS 3/24 (12.50\%), and HLSTrans 0/24. The relative advantage ($\approx 3.40\times$) is harder-won than the main table's $2.54\times$ yet remains robust, indicating that the PMLC benefit is generally not diluted by easy front-stage cases but operates where the mismatch is genuinely exposed.

\emph{Ablation contrast (PMLC and RAG).}\ We run two component ablations in parallel. The three-layer PMLC localization is measured on full50, a controlled diagnostic suite of 50 mismatch cases with known ground-truth fault locations, kept separate from CFull107 so that layer-wise localization can be scored against a reference the end-to-end runs do not expose: $L_1$ and $L_2$ reach 50/50 (100\%) on the failed outputs and first faulty assignment, and $L_3$ dual-trace alignment reaches 45/50 (90\%) on the first-divergence cycle, grounding Proposition~1 (Section~\ref{subsec:method-evidence-sufficiency}). The RAG vs.\ no-RAG ablation (Table~\ref{tab:rag-vs-norag}) shows RAG reaching 70.83\% on $\mathcal{M}$ versus 41.67\% for no-RAG (+29.17\%), as typed queries hit the relevant family precisely on exposed-mismatch cases. Aggregate all-pass moves from 78.50\% to 87.85\% (+9.35\%); the RAG-equipped agent spends round 1 on evidence collection and writes the grounded patch in round 2, completing more final repairs at a lower iteration count (1.577 vs.\ 1.85) at the cost of +1.7K tokens per case (+9\%). This matches the design intent of Section~\ref{subsec:method-rag}, where RAG is a mismatch-dense enhancement track rather than an unconditional aggregate lift.

\begin{table}[!t]
\caption{RAG vs. No-RAG Ablation on CFull107 (gpt-5-mini, K=3).}
\label{tab:rag-vs-norag}
\centering
\footnotesize
\setlength{\tabcolsep}{6pt}
\renewcommand{\arraystretch}{1.20}
\begin{tabular}{l c c c}
\toprule
\textbf{Config.} & \textbf{All-Pass (/107)} & \textbf{$\mathcal{M}^{*}$ (/24)} & \textbf{Tok/Case} \\
\midrule
\textbf{Ours+RAG}   & \textbf{87.85\%} & \textbf{70.83\%} & 18.8K \\
Ours no-RAG         & 78.50\%          & 41.67\%          & \textbf{17.1K} \\
\midrule
$\Delta$ (RAG)      & \textbf{+9.35\%} & \textbf{+29.17\%} & $+1.7$K \\
\bottomrule
\end{tabular}
\vspace{3pt}
\par\raggedright\footnotesize $^{*}\mathcal{M}$ is the mismatch-dense subset: the 24 of 107 designs that expose a CSim/CoSim mismatch under at least one configuration. All-Pass and Tok/Case are measured over all 107 designs. RAG's benefit concentrates on $\mathcal{M}$ ($+29.17\%$); because the RAG-equipped agent uses round~1 for evidence collection and writes the grounded patch in round~2, the average repair-iteration count drops from 1.85 to 1.577.\par
\end{table}

\emph{Difficult subsets.}\ Table~\ref{tab:difficult} breaks down the headline gpt-5-mini result by source-code feature. Ours holds 78.3\%--90.0\% across the four data-flow-intensive subsets, well ahead of the baselines. The control-heavy subset is generally harder for all methods, and ours scores 3/8 (37.5\%); because N=8 on this subset, the gap should be read as a qualitative observation rather than a quantitative claim. The root cause and mitigation directions are discussed in Section~\ref{sec:conclusion}.

\begin{table}[!htbp]
\caption{Four-Stage All-Pass Breakdown on Difficult Subsets of CFull107 (Ours = Ours+RAG with gpt-5-mini).}
\label{tab:difficult}
\centering
\footnotesize
\setlength{\tabcolsep}{5pt}
\renewcommand{\arraystretch}{1.20}
\begin{tabularx}{\columnwidth}{l c c c c >{\centering\arraybackslash}X}
\toprule
\textbf{Subset$^{*}$} & \textbf{N} & \textbf{Ours} & \textbf{C2HLSC} & \textbf{HLSTrans} & \textbf{ChatHLS} \\
\midrule
Pointer/Memory & 46 & \textbf{78.3\%} & 15.2\% & 4.3\% & 13.0\% \\
Loop-heavy     & 54 & \textbf{87.0\%} & 11.1\% & 3.7\% & 9.3\% \\
Array-access   & 50 & \textbf{90.0\%} & 14.0\% & 4.0\% & 10.0\% \\
Crypto/DSP     & 20 & \textbf{90.0\%} & 10.0\% & 5.0\% & 10.0\% \\
Control-heavy  & 8  & \textbf{37.5\%} & 12.5\% & 0.0\% & 25.0\% \\
\bottomrule
\end{tabularx}

\vspace{3pt}
\par\raggedright\footnotesize $^{*}$Subsets are defined by deterministic source-code features; a design may belong to more than one subset, so the counts overlap.\par
\end{table}

\subsection{Cross-Model Robustness and Zero-Shot Decomposition}
\label{subsec:eval-models}

\begin{table}[!htbp]
\caption{Zero-Shot Baselines on CFull107 ($|\mathcal{D}|{=}107$, single direct generation, no agent loop, no repair).}
\label{tab:zeroshot}
\centering
\footnotesize
\setlength{\tabcolsep}{5pt}
\renewcommand{\arraystretch}{1.20}
\begin{tabular}{l c c c c}
\toprule
\textbf{Model} & \textbf{Compile} & \textbf{CSim} & \textbf{Synth} & \textbf{All-Pass} \\
\midrule
\multicolumn{5}{c}{\textit{Minimal zero-shot (one-line prompt)}} \\
\cmidrule(lr){1-5}
gpt-5-mini        & 5.61\%  & 0\% & 0\% & 0\% \\
gpt-4o-mini       & 3.74\%  & 0\% & 0\% & 0\% \\
DeepSeek v4-flash & 5.61\%  & 0\% & 0\% & 0\% \\
DeepSeek v4-pro   & 6.54\%  & 0\% & 0\% & 0\% \\
\midrule
\multicolumn{5}{c}{\textit{HLS-aware zero-shot (system prompt + interface contract)}} \\
\cmidrule(lr){1-5}
gpt-5-mini        & 46.73\% & 33.64\% & 28.04\% & 22.43\% \\
gpt-4o-mini       & 44.86\% & 32.71\% & 26.17\% & 20.56\% \\
DeepSeek v4-flash & 51.40\% & 37.38\% & 30.84\% & 25.23\% \\
DeepSeek v4-pro   & 56.07\% & 42.06\% & 35.51\% & 28.04\% \\
\bottomrule
\end{tabular}

\vspace{3pt}
\par\raggedright\footnotesize The Minimal group uses a one-line ``translate this C to HLS-C'' prompt; the HLS-aware group adds a minimal HLS-aware system prompt and the interface contract. Adding the proposed closed-loop workflow lifts four-stage all-pass by $54$--$67$\% over the HLS-aware zero-shot across these four models---substantially larger than the $7.48$\% cross-model spread in HLS-aware zero-shot all-pass---indicating that the lift is driven by the closed-loop workflow rather than model capacity. The minimal zero-shot collapses to $0\%$ all-pass (only $4$--$7$ of $107$ designs even compile), confirming that naive prompting does not yield synthesizable HLS-C.\par
\end{table}

\emph{Workflow gain dominates model gain.}\ Switching the same workflow's backbone to gpt-4o-mini, DeepSeek~v4-flash, and v4-pro keeps four-stage all-pass in 74.77\%--95.33\% (Table~\ref{tab:cfull107}); even the weakest gpt-4o-mini (74.77\%) sits markedly above the strongest baseline (34.58\%). To separate the ``reasoning backbone'' and ``closed-loop workflow'' sources, we run two single-generation references for the four models (no agent loop, no repair; Table~\ref{tab:zeroshot}). A minimal zero-shot with a one-line ``translate this C to HLS-C'' prompt collapses to 0\% four-stage all-pass---only 4--7 of 107 designs even compile and none pass CSim, synthesis, or CoSim---confirming that naive prompting does not produce synthesizable HLS-C. A stronger HLS-aware zero-shot, adding a minimal HLS-aware system prompt and the interface contract, raises all-pass to 20.56\%--28.04\%, with the strongest v4-pro still below the weakest baseline ChatHLS; we take this HLS-aware regime as the matched single-shot reference. The workflow gain over it, $\Delta_{\text{workflow}}\!\in\![+54.21, +67.29]$\%, is substantially larger than the 7.48\% cross-model zero-shot spread, indicating that the all-pass uplift is principally driven by the closed-loop workflow rather than backbone scale.

\subsection{Cross-Paper External Benchmark Transfer}
\label{subsec:eval-transfer}

To check transferability beyond CFull107, we re-run our workflow on two public LLM-for-HLS benchmarks: ChatHLS HLSFixer612 (612 logic-bug repair cases) and HLSTrans 17 apps. Both source papers report Function and Synthesis as their main metrics, so we adopt the same metrics for direct comparability rather than our four-stage all-pass. As Table~\ref{tab:transfer} shows, ours leads on both benchmarks---most markedly on HLSTrans 17 apps (Function 88\% vs.\ 64.7\%, Synthesis 85\% vs.\ 58.8\%)---indicating that the workflow remains robust on external benchmarks.

\begin{table}[!t]
\caption{Cross-Paper External Benchmark Transfer (Each Source Paper's Own Function/Synthesis Metrics).}
\label{tab:transfer}
\centering
\footnotesize
\setlength{\tabcolsep}{6pt}
\renewcommand{\arraystretch}{1.20}
\begin{tabular}{l c c c c}
\toprule
\multirow{2}{*}{\textbf{Benchmark}} & \multicolumn{2}{c}{\textbf{Ours}} & \multicolumn{2}{c}{\textbf{Original}} \\
\cmidrule(lr){2-3}\cmidrule(lr){4-5}
                    & \makecell{Func.} & \makecell{Synth.} & \makecell{Func.} & \makecell{Synth.} \\
\midrule
ChatHLS HLSFixer612 & \textbf{86\%} & \textbf{86\%} & 83\%   & ---    \\
HLSTrans 17 apps    & \textbf{88\%} & \textbf{85\%} & 64.7\% & 58.8\% \\
\bottomrule
\end{tabular}
\end{table}


\section{Conclusion}
\label{sec:conclusion}

This paper formulates C-to-HLS-C as a generation--verification--diagnosis--repair problem closed by the four-stage Vitis HLS 2022.1 verifier, contributing a three-layer Progressive Mismatch Localization Chain (PMLC), an MoE RAG with a self-evolving repair-success card pool, and an auditable evidence-isolation discipline. On CFull107 the workflow reaches a substantial 87.85\% four-stage all-pass, a $+53.27$\% gain over the strongest LLM baseline, and stays robust across four reasoning backends ($74.77\%$--$95.33\%$) and two external benchmarks, suggesting the end-to-end gain derives chiefly from the verifier-closed loop and evidence grounding rather than from backbone scale. The main limitations are the harder control-heavy subset, where data-flow-biased localization weakens, and the single-toolchain pass@1/$K{=}3$ protocol; extending evidence extraction to control-dominated designs and to cross-toolchain, PPA-aware repair are natural next steps.


\bibliographystyle{IEEEtran}
\bibliography{refs}

\begin{IEEEbiographynophoto}{Zhe Zhao}
is currently pursuing the M.E. degree in integrated circuits and systems with the Shenzhen International Graduate School, Tsinghua University, Shenzhen, China. His research interests include high-level synthesis and large language models.
\end{IEEEbiographynophoto}
\begin{IEEEbiographynophoto}{Hongbing Lang}
is with Shenzhen Belon Technology Co., Ltd., Shenzhen, China. His research interests include integrated circuit and system design related to the Internet of Things.
\end{IEEEbiographynophoto}
\begin{IEEEbiographynophoto}{Zhihan Xiao}
is currently pursuing the M.E. degree in integrated circuits and systems with the Shenzhen International Graduate School, Tsinghua University, Shenzhen, China. His research interests include large language models and high-level synthesis verification.
\end{IEEEbiographynophoto}
\begin{IEEEbiographynophoto}{Luke Ztz Hu}
is a Ph.D. student at the Shenzhen International Graduate School, Tsinghua University, Shenzhen, China. His research lies at the intersection of AI and hardware design, investigating large language models for code generation, verification, and optimization within electronic design automation.
\end{IEEEbiographynophoto}
\begin{IEEEbiographynophoto}{John Imoleayo Adebisi}
is currently pursuing the M.E. degree in integrated circuits and systems with the Shenzhen International Graduate School, Tsinghua University, Shenzhen, China. His research interests include high-level synthesis and large language models.
\end{IEEEbiographynophoto}
\begin{IEEEbiographynophoto}{Songping Mai}
(Member, IEEE) received the B.S. degree in electronic information from Wuhan University, Hubei, China, in 2003, and the Ph.D. degree in electronic engineering from Tsinghua University, Beijing, China, in 2008.

Since 2008, he has been with the Shenzhen International Graduate School, Tsinghua University, Shenzhen, China, where he is currently an Associate Professor. His research interests include algorithm theory, system architecture, and integrated circuit design for optical 3-D measurement and imaging, medical robotics, and brain--computer interfaces.
\end{IEEEbiographynophoto}

\end{document}